\documentclass[a4paper,11pt]{article}
\pdfoutput=1
\usepackage{jheppub} 
\usepackage[utf8]{inputenc}       
\usepackage[T1]{fontenc}          
\usepackage{amsmath,amssymb}
\usepackage{graphicx}
\usepackage{xspace}
\usepackage{tikz}
\usepackage[font=small,labelfont=bf,format=plain,margin=0.05\textwidth]{caption}
\usepackage{listings}

\tikzstyle{arrow} = [draw, -latex, thick]
\tikzstyle{line} = [draw, -latex]
\usetikzlibrary{calc}
\usetikzlibrary{decorations.pathmorphing}	
\usetikzlibrary{decorations.markings}
\usetikzlibrary{arrows.meta}

\tikzset{
	fermion/.style={draw=black, postaction={decorate},
		decoration={markings,mark=at position .58 with {\arrow[draw=black,rotate=8]{Latex}}}},
	scalarnoarrow/.style={dashed,draw=black},
}

\title{\Large Addendum: Improved MSSM Higgs mass calculation using the 3-loop
  FlexibleEFTHiggs approach including $x_t$-resummation }
\author[a]{Thomas Kwasnitza,}
\author[a]{Dominik St\"ockinger,}
\author[b]{Alexander Voigt}

\affiliation[a]{Institut f\"ur Kern- und Teilchenphysik,
  TU Dresden,\\ Zellescher Weg 19, 01069 Dresden, Germany}

\affiliation[b]{Institute of Energy and Life Science, Flensburg University of Applied Sciences,\\
  Kanzleistraße 91--93, 24943 Flensburg, Germany}

\emailAdd{thomas.kwasnitza@mailbox.tu-dresden.de}
\emailAdd{dominik.stoeckinger@tu-dresden.de}
\emailAdd{alexander.voigt@hs-flensburg.de}

\abstract{ In this addendum we present the stand-alone C++ program
  \prog, which implements the 3-loop \feft\ approach to calculate the
  lightest $CP$-even Higgs boson pole mass in the real MSSM at \NCLL\
  and \NCLO\ with $x_q$ resummation, presented in ref.~\cite{Kwasnitza:2020wli}.}

\arxivnumber{}

\hypersetup{
  pdftitle    = {Improved MSSM Higgs mass calculation using the 3-loop FlexibleEFTHiggs approach including xt-resummation},
  pdfauthor   = {Thomas Kwasnitza, Dominik Stöckinger, Alexander Voigt},
  pdfkeywords = {MSSM, Higgs, Mass}
}

\allowdisplaybreaks

\newcommand{\fseft}{\texttt{MSSMEFTHiggs3L}\@\xspace}

\newcommand{\feft}{Flex\-ib\-le\-EFT\-Higgs}
\newcommand{\prog}{\texttt{MSSMEFTHiggs3L}}

\newcommand{\code}[1]{\lstinline|#1|}  %

\newcommand{\pole}{\text{pole}}

\newcommand{\MS}{\ensuremath{M_S}\xspace}

\newcommand{\Qmatch}{\ensuremath{Q_\text{match}}}

\newcommand{\NCLO}{\ensuremath{\text{N}^3\text{LO}}\xspace}
\newcommand{\NCLL}{\ensuremath{\text{N}^3\text{LL}}\xspace}

\newcommand{\figref}[1]{figure~\ref{#1}}

\newcommand{\secref}[1]{section~\ref{#1}}

\newcommand{\Qpole}{\ensuremath{Q_\pole}}
\newcommand{\Qlow}{\ensuremath{Q_{\text{low}}}}

\newcommand{\FS}{\texttt{FlexibleSUSY}\xspace}

\toccontinuoustrue 

\begin{document}
\maketitle
\newpage

\section{Introduction}

We present the C++ program \prog, which implements the 3-loop \feft\
state-of-the art calculation of $M_h$ in the real MSSM at \NCLL\ and
\NCLO\ with $x_q$ resummation.  The program is based on the \FS\ model
\texttt{NUHMSSMNoFVHimalaya} and implements the matching and running
described in our original publication, thus reproducing the results
presented there. The program provides an easy-to-use SLHA interface
for the MSSM input parameters and prints the value of $M_h$ as a
single number to \texttt{stdout}.

We have structured the addendum as follows. In \secref{sec:building}
we describe the technical details relevant for building the
program. In \secref{sec:interface} we discuss the user interface
and relevant configuration options.  Finally, we comment on the
upcoming integration of the refined FlexibleEFTHiggs approach with
full-model parametrization into the general \FS\ package.

\section{Installation and usage of the stand-alone code}
\label{sec:building}

The \prog\ program can be downloaded as compressed package from
\begin{center}
  \footnotesize\url{https://flexiblesusy.hepforge.org/downloads/FlexibleEFTHiggs/MSSMEFTHiggs3L.tar.gz}
\end{center}
To build \prog, the boost C++ library, the Eigen3 library, the GNU
Scientific Library and the Himalaya library~\cite{Harlander:2017kuc} (version 4.0.0 or
higher) are required. For installation instructions of the Himalaya
library see e.\,g.\ ref.~\cite{Harlander:2017kuc}.

After the package has been extracted, it can be configured and
compiled by running the following commands:
\begin{lstlisting}
$ ./configure --enable-himalaya --enable-fflite \
     --with-himalaya-incdir=${HIMALAYA_DIR}/include \
     --with-himalaya-libdir=${HIMALAYA_DIR}/build \
     --with-models=NUHMSSMNoFVHimalaya
$ make
\end{lstlisting}
The variable \texttt{HIMALAYA\_DIR} contains the path to
Himalaya root directory, required for the 3-loop pole-mass matching.
Due to an improved numerical robustness, we recommend the
configuration with the shipped 1-loop integral library
\texttt{FFLite}. For more options see \texttt{./configure -h}.  After
the compilation has finished, the program can be run with the shipped SLHA
input file as follows:
\begin{lstlisting}
$ SLHA_INPUT=models/NUHMSSMNoFVHimalaya/LesHouches.in.NUHMSSMNoFVHimalaya
$ models/NUHMSSMNoFVHimalaya/run_NUHMSSMNoFV_fefthiggs.x \
     --slha-input-file=$SLHA_INPUT
\end{lstlisting}
Running the program with the shipped SLHA input file yields the
following output for the lightest $CP$-even Higgs pole mass $M_h$ on
command line:
\begin{lstlisting}
123.522878
\end{lstlisting}

\section{Interface and configuration options}
\label{sec:interface}

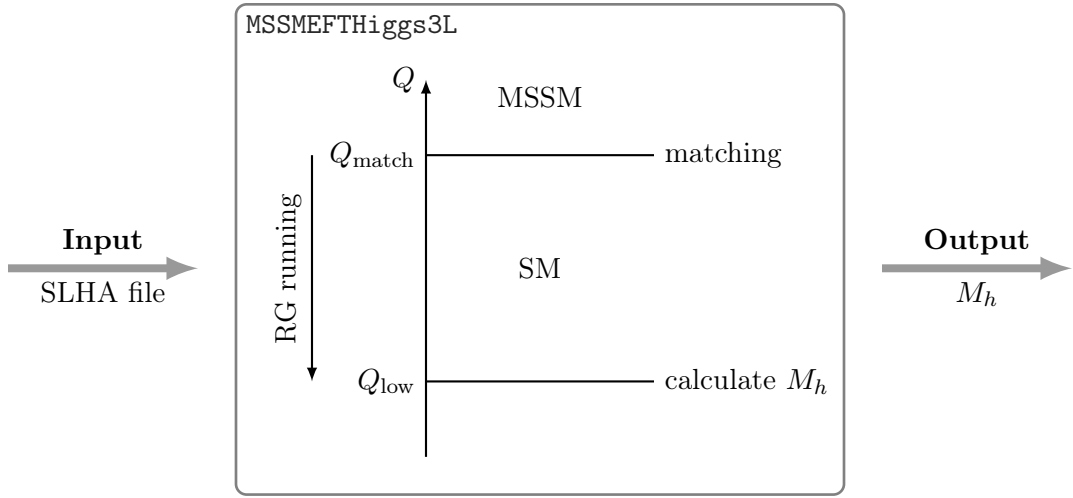
\begin{figure}[tb]
	\centering
	\begin{tikzpicture}
	\path [line,line width=3pt, draw=black!40] (-5.5,2.5) --node[above] {\textbf{Input}} node[below] {SLHA file}  (-3,2.5);
	\path [line,line width=3pt, draw=black!40 ]
	(6,2.5)-- node[above] {\textbf{Output}} node[below] {$M_h$}  (8.5,2.5);
	\begin{scope}
	\path[arrow] (0,0) -- (0,5) node[left]{$Q$};
	\draw[thick] (0,4) node[left]{$\Qmatch$} -- node[above = 0.5cm]{MSSM} (3,4) node[right]{matching};
	\draw[thick] (0,1) node[left]{$\Qlow$} -- (3,1) node[right]{calculate $M_h$};
	\draw[thick] (1.5,2.5) node[]{SM};
	\path[arrow,latex-] (-1.5,1) -- node[above,rotate=90]{RG running} (-1.5,4);
	\draw[line width=1, draw=gray,rounded corners] (-2.5,6) rectangle (5.5,-.5);
	\node[color=black!85, anchor=west] at (-2.5,5.7)  {\fseft};
	\end{scope}
	\end{tikzpicture}
	\caption{Interface of the stand-alone code}
	\label{fig:interface}
\end{figure}

The \prog\ program expects the MSSM input parameters in SLHA-1 format,
see \figref{fig:interface}. It calculates the lightest $CP$-even Higgs
boson pole mass $M_h$ in the real MSSM with fermion and sfermion
flavour conservation and with the non-universal Higgs mass parameters
$m_{H_u}^2$ and $m_{H_d}^2$ fixed by the electroweak symmetry breaking
conditions, as described in the original publication.  When the
calculation has finished successfully, the program writes the decimal
floating-point value of the $M_h$ to \texttt{stdout}.

The multi-loop contributions entering the Higgs mass calculation are
controlled by the configuration options in the \FS\ block of the SLHA
input.  A detailed documentation of the flags is given in ref.~\cite{Athron:2017fvs}.
Here, we discuss the relevant options in the \FS\ block of the SLHA
input, which controls the individual corrections of the Higgs pole
mass calculation. Depending on the desired precision of the Higgs pole
mass calculation, we present two configurations.

\paragraph{Default 3-loop precision (i.\,e.\ $M_h$ at \NCLO, \NCLL and with $x_q$-resummation):}
For a consistent \feft~calculation at this order, the following relevant flags have to be set in the SLHA input:
\begin{lstlisting}
Block FlexibleSUSY
 4   3           # pole mass loop order
 5   3           # EWSB loop order
 6   4           # beta-functions loop order
 7   3           # threshold corrections loop order
 8   1           # Higgs 2-loop corrections O(alpha_t alpha_s)
 9   1           # Higgs 2-loop corrections O(alpha_b alpha_s)
10   1           # Higgs 2-loop  O((alpha_t + alpha_b)^2)
11   1           # Higgs 2-loop corrections O(alpha_tau^2)
13   2           # Top pole mass QCD corrections (1 = 2L, 2 = 3L)
18   0           # pole mass scale in the EFT (0 = Mt)) 
19   0           # EFT matching scale (0 = SUSY scale)
20   2           # EFT loop order for yt matching
21   3           # EFT loop order for lambda matching
24   124111321   # individual threshold correction loop orders
26   1           # Higgs 3-loop corrections O(alpha_t alpha_s^2)
27   0           # Higgs 3-loop corrections O(alpha_b alpha_s^2)
28   0           # Higgs 3-loop corrections O(alpha_t^2 alpha_s)
29   0           # Higgs 3-loop corrections O(alpha_t^3)
30   0           # Higgs 4-loop corrections O(alpha_t alpha_s^3)
\end{lstlisting}
The meaning of each flag is described in the associated comment. The
user should be aware that deviations from the displayed flag
configuration usually result in a reduced precision of the
calculation.  In the following we briefly describe a selection of
adjustments:
\begin{itemize}
\item \textbf{Flag 18} This flag can be used to set the
  renormalization scale $\Qpole$ (in GeV), at which the Higgs pole
  mass $M_h$ is calculated in the SM. Possible values are $\Qpole=0$,
  which corresponds to $\Qpole=M_t$, or any positive value $\Qpole>0$.
  This flag can be used to vary the renormalization scale in order to
  estimate the low-scale uncertainty as described in sec.~8.3.2.

\item \textbf{Flag 19} This flag can be used to set the matching scale
  $\Qmatch$ (in GeV) at which $\lambda$ is determined. Possible values
  are $\Qmatch=0$, which corresponds to $\Qmatch=\MS$, or any positive
  value $\Qmatch>0$.
  This flag can be used to vary the matching scale in order to
  estimate the high-scale uncertainty as described in the vicinity of
  eq.~(8.11).

\item \textbf{Flag 20} This flag has a different meaning than
  described in the documentation in ref.~[56], where it controls the
  loop order of the upwards matching from the SM to the full model.
  Our calculation does not require any upwards matching and we use it
  to control the downwards matching of SM-like gauge and Yukawa
  couplings.  Possible values are 0 (tree-level), 1 (1-loop) and 2
  (2-loop).  For a calculation of $M_h$ at \NCLO\ and \NCLL, the flag
  must be set to $2$. Reducing the value to 1 or 0 reduces the
  large-log resummation to NNLL or LL, respectively.

\item \textbf{Flag 21} This flag controls the loop order for the
  calculation of $\lambda$ and specifies the contributions in
  eq.~(4.28a). Possible values are 0 (tree-level), 1 (1-loop), 2
  (2-loop) and 3 (3-loop).  For a calculation of $M_h$ at \NCLO\ and
  \NCLL, the flag must be set to $3$.  If numerical instabilities
  occur, it may be beneficial to reduce the loop order of the
  calculation of $\lambda$ to 2-loop (gauge-less limit) and therefore
  restrict the precision to NNLL and NNLO.
\end{itemize}

\paragraph{Minimal 2-loop precision (i.\,e.\ $M_h$ at NNLO, NNLL and with $x_q$-resummation):}
The minimal flag configuration to achieve a \feft~calculation at this
precision requires the following configuration settings in the SLHA
input:
\begin{lstlisting}
Block FlexibleSUSY
 4   2           # pole mass loop order
 5   2           # EWSB loop order
 6   3           # beta-functions loop order
 7   2           # threshold corrections loop order
 8   1           # Higgs 2-loop corrections O(alpha_t alpha_s)
 9   1           # Higgs 2-loop corrections O(alpha_b alpha_s)
10   1           # Higgs 2-loop  O((alpha_t + alpha_b)^2)
11   1           # Higgs 2-loop corrections O(alpha_tau^2)
13   1           # Top pole mass QCD corrections (1 = 2L, 2 = 3L)
18   0           # pole mass scale in the EFT (0 = Mt)) 
19   0           # EFT matching scale (0 = SUSY scale)
20   1           # EFT loop order for yt matching
21   2           # EFT loop order for lambda matching
24   112111111   # individual threshold correction loop orders
26   0           # Higgs 3-loop corrections O(alpha_t alpha_s^2)
27   0           # Higgs 3-loop corrections O(alpha_b alpha_s^2)
28   0           # Higgs 3-loop corrections O(alpha_t^2 alpha_s)
29   0           # Higgs 3-loop corrections O(alpha_t^3)
30   0           # Higgs 4-loop corrections O(alpha_t alpha_s^3)
\end{lstlisting}

\section{Outlook}

In this addendum, we have presented the stand-alone program \prog,
which has been developed for the Higgs mass calculation presented in
the original publication.

We plan to implement the refined FlexibleEFTHiggs approach with
full-model parametrization into the general \FS\ package. This allows
to apply the calculation to models beyond the real MSSM, such as the
NMSSM etc.  The planned integrated version will also allow access to
the full pole-mass spectrum of the model as well as the computation of
other observables.

\bibliographystyle{JHEP}
\bibliography{flexibleefthiggs_addendum}

\providecommand{\href}[2]{#2}\begingroup\raggedright\begin{thebibliography}{1}

\bibitem{Kwasnitza:2020wli}
T.~Kwasnitza, D.~St\"ockinger and A.~Voigt, \emph{{Improved MSSM Higgs mass
  calculation using the 3-loop FlexibleEFTHiggs approach including
  $x_{t}$-resummation}},
  \href{http://dx.doi.org/10.1007/JHEP07(2020)197}{\emph{JHEP} \textbf{07}
  (2020) 197}, [\href{http://arxiv.org/abs/2003.04639}{{\texttt 2003.04639}}].

\bibitem{Harlander:2017kuc}
R.~V. Harlander, J.~Klappert and A.~Voigt, \emph{{Higgs mass prediction in the
  MSSM at three-loop level in a pure $\overline{{\text {DR}}}$ context}},
  \href{http://dx.doi.org/10.1140/epjc/s10052-017-5368-6}{\emph{Eur. Phys. J.}
  \textbf{C77} (2017) 814}, [\href{http://arxiv.org/abs/1708.05720}{{\texttt
  1708.05720}}].

\bibitem{Athron:2017fvs}
P.~Athron, M.~Bach, D.~Harries, T.~Kwasnitza, J.-h. Park, D.~Stöckinger
  et~al., \emph{{FlexibleSUSY 2.0: Extensions to investigate the phenomenology
  of SUSY and non-SUSY models}},
  \href{http://dx.doi.org/10.1016/j.cpc.2018.04.016}{\emph{Comput. Phys.
  Commun.} \textbf{230} (2018) 145--217},
  [\href{http://arxiv.org/abs/1710.03760}{{\texttt 1710.03760}}].

\end{thebibliography}\endgroup

\end{document}